# Investigating the impact of BTI, HCI and time-zero variability on neuromorphic spike event generation circuits

Shaik Jani Babu[1], Rohit Singh[1], Siona Menezes Picardo[2], Nilesh Goel[2], and Sonal Singhal[1*]

[1]Electrical Engineering, Shiv Nadar University, Greater Noida, India
[2]Electronics and Electrical Engineering, BITS Pilani Dubai Campus, Dubai, UAE

**ABSTRACT**

Neuromorphic computing refers to brain-inspired computers, that differentiate it from von Neumann architecture. Analog VLSI based neuromorphic circuits is a current research interest. Two simpler spiking integrate and fire neuron model namely axon-Hillock (AH) and voltage integrate, and fire (VIF) circuits are commonly used for generating spike events. This paper discusses the impact of reliability issues like Bias Temperature instability (BTI) and Hot Carrier Injection (HCI), and time-zero variability on these CMOS based neuromorphic circuits. AH and VIF circuits are implemented using HKMG based 45nm technology. For reliability analysis, industry standard Cadence RelXpert tool is used. For time-zero variability analysis, 1000 Monte-Carlo simulations are performed.

## 1. INTRODUCTION

Neuromorphic systems are based on biologically-inspired neural cells that communicate through a discrete-amplitude, continuous time mechanism. Implementation of spiking neurons is extremely useful for a variety of applications like real-time low power large-scale neural systems, brain-machine interface [1]. Several models have been proposed in the past that aid in investigating the behavior of neuronal systems mathematically while encompassing essential characteristics of neural processing.[2] They are mainly categorized into biophysically realistic, complex conductance-based models namely the Hodgkin-Huxley (H-H) and less biologically realistic Integrate-and-Fire (I&F) model. I&F models can be implemented by Silicon Neuron Circuits (SiN) are analog/digital very large-scale integration (VLSI) circuits that replicate the electro-physiological performance of biological neurons and their conductance [1]. The work in this paper involves implementation of circuits based on the electronic spiking neuron model and I&F model namely the Axon Hillock circuit and a Voltage Amplifier I&F circuit respectively. The Axon-Hillock circuit proposed by Mead [3] consists of a membrane capacitance that models the biological membrane of the neuron and an amplifier block with two series inverting amplifiers with a feedback capacitor. An action potential (spike) is generated when the voltage on the integrating capacitor exceeds the switching threshold of the first inverter in the amplifier. The Voltage Amplifier I&F (VIF) circuit, alternatively, uses a basic transconductance amplifier to compare the membrane voltage with an adjustable threshold voltage. This circuit additionally produces explicit refractory periods following spike generation, which is an essential feature for neuron communication [4,1]. Investigation of the impact of reliability issues such as Bias Temperature Instability (BTI) and Hot Carrier Injection (HCI), and time-zero variability on the stated CMOS-based neuromorphic circuits are carried out. The reliability issues and time-zero variability lead to degradation of threshold voltages of the transistors [5-9]; involved in the generation of spiking event in these circuits. The frequency of spiking events is the most critical parameter in neuromorphic circuits as they are known to be information carriers in biological neurons. This aspect of the spiking event is primarily considered to study the bearing of BTI and HCI. The circuits discussed in this paper are implemented using HKMG based 45nm technology and the reliability analysis is carried out using the industry standard Cadence RelXpert tool. Monte-Carlo simulations are performed to analyze the time-zero variability. The paper is organized as follows: section 2 explains about two neuromorphic circuits used in this work. Section 3 outlines an approach for the efficient reliability analysis. Section 4 presents results and discussion. Finally, section 5 provides conclusion of this paper.

## 2. NEUROMORPHIC SPIKE EVENT GENERATION CIRCUIT

In this section two commonly used neuromorphic spike event generation (i) Axon-Hillock (AH) (ii) Voltage Integrate & Fire (VIF) circuits are discussed.

## 2.1. Axon - Hillock (AH) circuit

The spiking neuron circuit namely Axon-Hillock (AH) is shown in figure 1(a). A spike event is generated when membrane potential reaches a set threshold voltage. Here current source ($I_{inj}$) model the synaptic current and charges the membrane capacitor ($C_M$) and develops membrane voltage ($V_{MEM}$). $V_{MEM}$ continues to increase till it reaches the switching threshold voltage set by an inverter (PM0-NM0). This results in the spike event generation ($V_{SPK}$) at output. Generated Spike leads to steep charging of $C_M$ through positive feedback capacitor ($C_F$) and also provides a current reset path through transistor NM2. Reset path is additionally controlled by an external voltage $V_{CK}$. Fig. 1(b) shows simulated waveform of $V_{MEM}$ and $V_{SPK}$ voltages. Spike event generation of AH circuit is shown in the inset of Fig. 1(b).

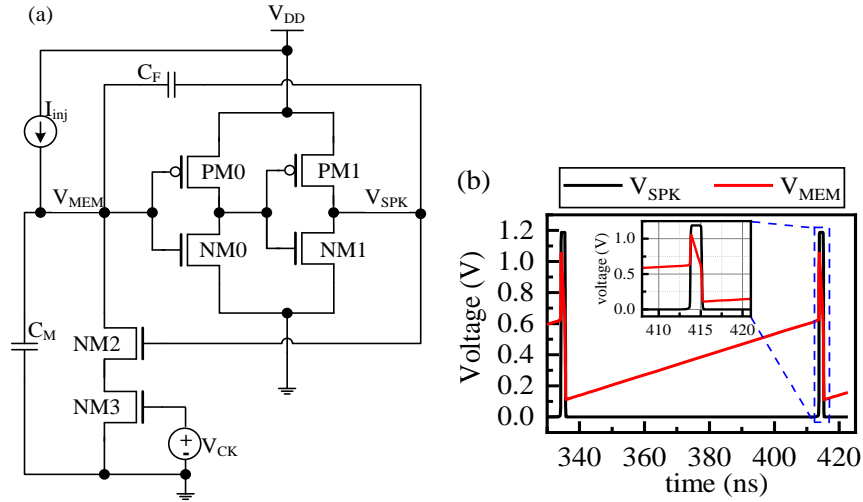

**Fig. 1:** (a) Schematic of Axon Hillock (AH) circuit (b) Simulated waveform of $V_{MEM}$ and $V_{SPK}$

## 2.2. Voltage Integrate & Fire (VIF) circuit

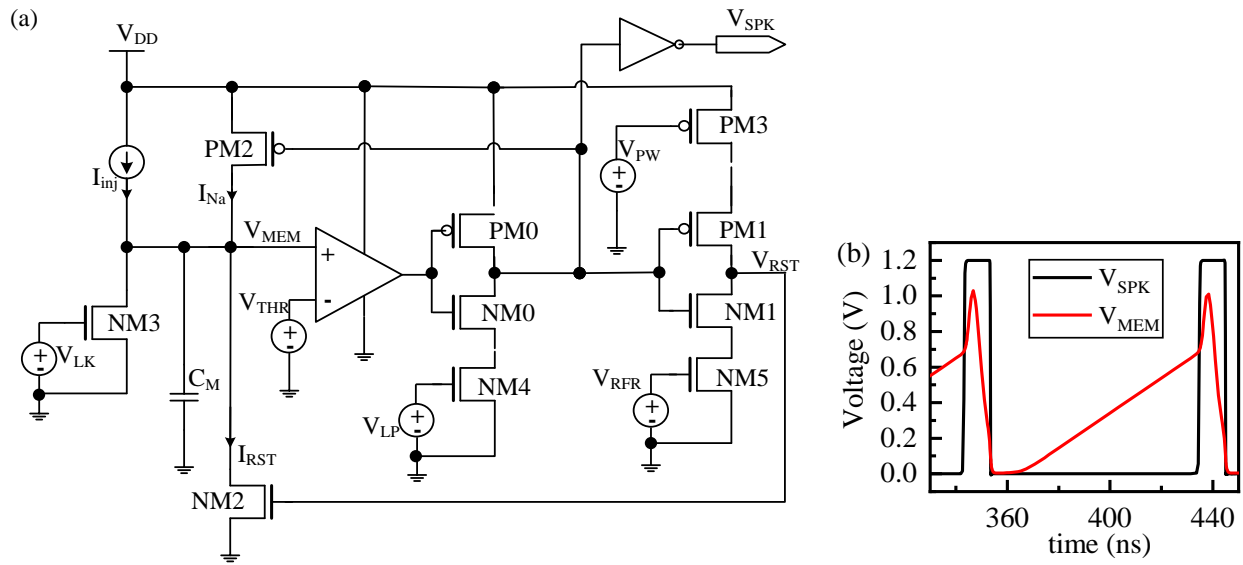

**Fig. 2:** (a) Schematic of Voltage Integrate and Fire (VIF) circuit (b) Simulated waveform of $V_{MEM}$ and $V_{SPK}$ voltages

Fig. 2(a) shows, the Voltage Integrate and Fire (VIF) spiking neuron Circuit. In VIF circuit, spiking behaviour (set threshold, refractory period) is governed using external voltage source. Current source ($I_{inj}$) models the synaptic current and charges the membrane capacitor ($C_M$) and thus develops membrane potential ($V_{MEM}$). A difference amplifier, configured as comparator, is used to compare the membrane potential ($V_{MEM}$) with set threshold voltage ($V_{THR}$). Comparator output goes high ($V_{DD}$) as $V_{MEM}$ reaches above $V_{THR}$. This results in steep charging of $V_{MEM}$ by sodium current ($I_{Na}$) through PM2. The pulse width of spike is controlled using transistor PM3 and external voltage source ($V_{PW}$). During $V_{RST}$ high, NM2 transistor also gets turned ON and provides a reset current ($I_{RST}$) a discharge path for membrane capacitor. The condition for $V_{MEM}$ to be reset is $I_{RST}$ should be greater than sum of $I_{inj}$ and $I_{Na}$. The refractory period is governed by transistor NM5 which is kept in sub threshold region using external voltage source ($V_{RFR}$). In no synaptic current event, membrane capacitor discharges through sub-threshold operated transistor NM3. Fig. 2(b) shows simulated waveform of $V_{MEM}$ and $V_{SPK}$ voltages.

## 3. METHODOLOGY

Fig. 3 shows the flow chart of the simulation framework to evaluate the impact of BTI, HCI and process variability. Technology library from foundry is used to create AH and VIF circuit netlist. Industry standard reliability simulator ©Cadence Relxpert tool is used to retrieve the aged/degraded circuit netlist. The aged and fresh circuit netlists are then given to SPICE simulator for aged and fresh simulation for reliability analysis. MC SPICE simulations of 1000 runs are performed to evaluate the process variability for both AH and VIF netlist. AH and VIF performance metrics are extracted in the MATLAB environment. Since the frequency of spiking events is the most critical parameter of AH and VIF circuit, simulations are carried out to observe the spiking frequency by varying synaptic current ($I_{inj}$). Synaptic current is varied in range of 0.2μA to 60μA in steps of 20. Simulation is run for fresh and aged condition with a considered lifetime span of 10 years. Performance of spiking neuron circuit is characterized by calculating the percentage deviation in spiking frequency of aged and fresh circuit. Percentage deviation is calculated by using equation (1):

$$\text{Percentage deviation in } F_{SPK} = \frac{(\text{aged\_F}_{SPK} - \text{fresh\_F}_{SPK})}{\text{fresh\_F}_{SPK}} * 100 \quad (1)$$

Where fresh\_$F_{SPK}$ and aged\_$F_{SPK}$ are spiking frequency values at time-zero and after degradation respectively.

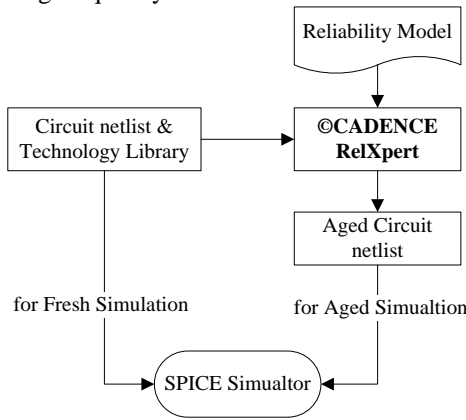

**Fig. 3:** Simulation flow for reliability analysis

## 4. RESULTS AND DISCUSSIONS

This section presents the results of fresh and aged spiking frequency of AH and VIF circuit. Process variability for the neuron are also discussed. Probit plots are shown and discussed to analyze the process variability.

Fig. 4(a) shows the percentage deviation in spiking frequency due to combined effect of BTI and HCI on both AH and VIF circuits. It is observed that spiking frequency of AH neuron circuit does not vary significantly (~ 2%). In AH circuit, transistor NM2 is placed in the reset path and is largely responsible in governing the frequency of spike event. It is observed from Fig. 1(b), that transistor NM2 remains in OFF state for 99% of its total time period and is therefore insensitive to degradation. Remaining transistors (NM0, PM0, NM1 and PM1) suffers from BTI and HCI degradation, and affect the switching threshold voltage of $V_{MEM}$. However, degradation in these transistors do not impact the frequency of spike event. This in turn makes spiking frequency of AH circuit insensitive to BTI and HCI degradation. It is observed that spiking frequency of VIF neuron circuit degraded significantly (~ 12%) till 2μA of synaptic current ($I_{inj}$) and further increase of $I_{inj}$, VIF circuit does not generate spike. The spike event generation is largely governed by transistors NM2 and PM2, which are placed in reset and positive feedback path respectively as discussed in section 1. As transistor NM2 is stressed for (5-10) % of its total time period and is therefore degraded. This result lesser $I_{RST}$ and hence a slower discharge path for membrane capacitor leading to degradation of spiking frequency is observed. With increase in synaptic current (>2μA), spiking frequency will increase, resulting in more stress time for transistor NM2. Hence more shift in threshold voltage of NM2 is observed. Degraded reset current is unable to discharge the membrane capacitor which result in no spike event generation and is observed in Fig. 4 (a). Fig. 4(b) shows spiking frequency ($F_{SPK}$) distribution due to process variability for AH and VIF neuron circuits. Variation of $F_{SPK}$ for AH neuron circuit is found negligible having a standard deviation. Whereas for VIF, it is observed to be significant value. $F_{SPK}$ distribution for AH follows gaussian distribution, whereas for VIF also follows gaussian distribution having a marginal deviation at tail ends.

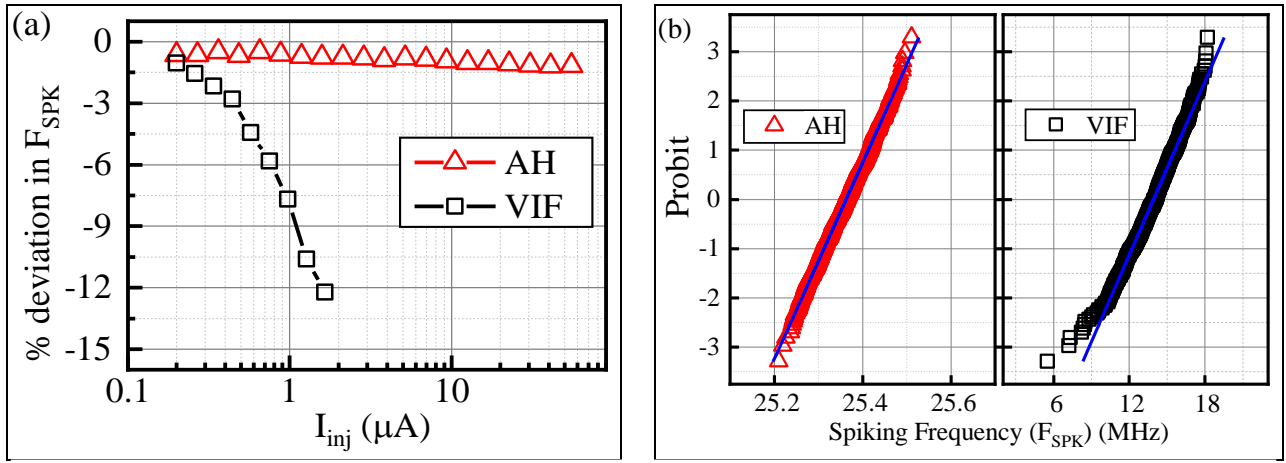

**Fig. 4:** (a) Percentage deviation in spiking frequency ($F_{SPK}$) of AH and VIF circuits (b) Spike frequency ($F_{SPK}$) distribution for AH and VIF circuits (blue line shows the expected gaussian distribution)

## 5. CONCLUSIONS

This paper studies the impact of reliability issues like Bias Temperature instability (BTI) and Hot Carrier Injection (HCI) on neuromorphic circuits namely Axon Hillock and Voltage Integrate and Fire. It also investigates process variability at time-zero of spike event neuron circuits. Spiking frequency ($F_{SPK}$) is the most critical parameter for a neuron circuit, hence it is used to investigate the impact of BTI and HCI on these neuron circuits. $F_{SPK}$ degradation for AH circuit due to the impact of both HCI and BTI is observed around 2%. While for VIF circuit, with increase of synaptic current a larger deviation in frequency is observed and eventually no spike event generation occurs at high synaptic current. It is due to degradation in NM2 transistor which plays major role in reset path. AH circuit is found to be insensitive to time-zero variability while VIF circuit shows a marginal deviation at tail ends along with large variation in spiking frequency. The present work provides an opportunity to investigate and build a reliability aware design of the considered neuromorphic circuits.

**Acknowledgement**

The authors are grateful to the Electrical Engineering department of Shiv Nadar University for providing simulation tools and computational resources which are used in this work. Authors are also thankful to the Electrical and Electronics department of BITS Pilani Dubai campus.